\documentstyle[sprocl,epsfig]{article}

\def\Journal#1#2#3#4{{#1} {\bf #2}, #3 (#4)}

\def\NPB{{\em Nucl. Phys.} B}
\def\PLB{{\em Phys. Lett.}  B}
\def\PRL{\em Phys. Rev. Lett.}
\def\PRD{{\em Phys. Rev.} D}
\def\ZPC{{\em Z. Phys.} C}

\def\be{\begin{equation}}
\def\ee{\end{equation}}
\def\bea{\begin{eqnarray}}
\def\eea{\end{eqnarray}}

\begin{document}

\vspace*{-2cm}
\begin{flushright}
MC-TH-97/11
\end{flushright}

\title{Photon '97: Theory Summary\footnote{Talk presented at the International
Conference on the Structure and Interactions of the Photon, including
the 11th International Workshop of Photon-Photon Collisions, Egmond aan
Zee, The Netherlands, May 1997.}}

\author{ Jeffrey R. Forshaw }

\address{Department of Physics \& Astronomy, University of Manchester,\\
Manchester. M13 9PL. UK.}

\maketitle\abstracts{Some recent developments in the physics of photon induced
reactions are discussed. My presentation is biased towards
HERA physics with David Miller's talk being biased towards the 
$\gamma \gamma$ topics \cite{DJM}. Within the context of the data which
were presented, I shall concentrate upon the the following topics:
diffraction; jets; prompt photons; open charm and charmonium.}

\section{Diffraction}
I'm going to restrict myself to diffractive phenomena in $\gamma p$
interactions (where the photon can be real or virtual).
For the present purposes ``diffraction'' means that there is a
rapidity gap in the final state (I'll have more to say on this later).
Let's start by recalling some results on total rates \cite{Arneodo}. 
In particular, I want
to discuss the $W$-dependence of total rates ($W$ is the $\gamma p$
invariant mass). A convenient way to parameterize the data on the
$W$-dependence is to extract an effective pomeron intercept,
$\alpha_P(0)$. It is to be understood that this value would be the true
pomeron intercept if the physics were solely due to exchange of a single
Regge pole. Recall that total hadronic cross-sections and exclusive
photoproduction of light vector mesons (e.g. $\gamma p \to V p$
where $V$ is a vector meson and the photon is close to its mass-shell,
i.e. $Q^2 \approx 0$) can be described with a single value of
$\alpha_P(0) \approx 1.08$ \cite{DonLan}. However, there are processes
which do not follow this trend. The dissociation process, 
$\gamma^* p \to X p$ ($X$ denotes the dissociation products which are
distant in rapidity from the outgoing proton), is characterized by   
a rather large effective intercept, i.e. $\alpha_P(0) = 1.18 \pm 0.02
\pm 0.04$ \cite{Ebert}. There is also a tendency for the intercept to
grow as $Q^2$ increases \cite{Piotrzkowski}. In addition, the growth for
$\gamma p \to J/\psi \; p$ is larger still, $\alpha_P(0) \approx 1.3$ (even
for photoproduction, i.e. $Q^2 \approx 0$). The situation for light meson
production off virtual photons is less clear \cite{Kohne,Crittenden}.
There is a suggestion that the $W$ dependence 
is much steeper than it would be for
on-shell photons, and that it steepens as $Q^2$ rises
\cite{Crittenden}. However the conclusion relies upon extrapolating from
low energy data, where the recent E665 measurement \cite{E665} (of a high
cross-section value at $W \approx 11$ GeV) confuses the issue. Data from
Hermes should help sort this out \cite{Kolstein}. It seems that a large
photon virtuality or a large quark mass is correlated with a more
rapidly rising cross-section (recall also that the deep inelastic total
cross-section rises rapidly with increasing $W$, i.e. at small
$x$). Michele Arneodo asked ``just what is the
scale which determines how steep the rise is?'' \cite{Arneodo}. 
To gain some insight into the physics which determines the answer to
this question is the next part of my talk. 
 
\subsection{A physical picture of diffraction}

Consider shining a coherent beam of partons onto a target at rest
and let $z_i$ \& $b_i$ be the fraction of the total beam
energy carried by parton $i$ and its position in the plane transverse to
its direction of travel respectively. We expect that, as the beam energy
increases, so too does the probability that the parton passes through
the target undeflected (i.e. any momentum transfer it receives is too
small to deflect it appreciably). Of course the target itself can be
broken up by the momentum transfer (or scattered into some excited state
or scattered elastically). This type of event can have
a big rapidity gap between the final state partons and the products of
the target. It also follows that the eigenstates of this kind of
reaction will be states of fixed $z_i$ and $b_i$. The target simply
alters the profile of the incoming beam. The coherent sum over the final
state partons will lead to a state which has some overlap with the
initial state (elastic scattering) and also will lead to states which
have non-zero overlap with other final states. The analogy with optical
diffraction is clear (the parton states play the analogous role
the the Huygens wavelets) and hence the name. It's now time to turn to
diffraction in photon induced reactions.

To describe the hadronic interactions of the photon we need to consider
its fluctuation into a $q \bar{q}$ pair and the subsequent interaction
with the target (at high beam energy, the pair will typically be
produced way upstream of the target). We describe the $q \bar{q}$
fluctuations with the wavefunction, $\psi_{\gamma}(z,r)$, i.e.
\begin{equation}
|\gamma \rangle = \int dz d^2r \; \psi_{\gamma}(z,r)|z,r \rangle.
\end{equation}
The defining statement that the eigenstates of diffraction are
states of fixed $z$ and $r$ can be written
\begin{equation}
\hat{T}|z,r \rangle = i \tau(s,b;z,r) |z,r \rangle
\end{equation}
where $\hat{T}$ is the operator which determines how the $q \bar{q}$
state scatters off the target and the eigenvalue $i \tau(s,b;z,r)$ is the 
associated amplitude for scattering the partons elastically (I just
chose to take out a factor of $i$ since it turns out that the amplitude
is dominanted by its imaginary part, i.e. $\tau$ is real). 
The $\gamma-$Target invariant mass
is denoted $s=W^2$ and $b$ is the impact parameter for the collision.

The {\bf elastic scattering} amplitude, 
\begin{equation}
A^{{\rm el}}(s,t) = \int d^2b \; {\rm e}^{i q \cdot b} \langle \gamma|
\hat{T} | \gamma \rangle,
\end{equation}
(we have taken the Fourier transform so as to get the amplitude in terms
of the momentum transfer, $-t = q^2 > 0$) can then be written:
\begin{eqnarray}
\frac{{\rm Im} A^{{\rm el}}(s,t=0)}{s} &=& \int dz d^2r \;
|\psi_{\gamma}(z,r)|^2 \sigma(s,r) \nonumber \\
&=& \sigma^{\gamma T}_{{\rm tot}}(s),
\end{eqnarray}
where $\sigma^{\gamma T}_{{\rm tot}}(s)$ is the total $\gamma-$Target
cross-section and
$$ \int d^2b \frac{\tau(s,b;z,r)}{s} \equiv \sigma(s,r) $$
is the total cross-section for scattering the $q \bar{q}$ pair off the
target (for convenience we suppress any dependence on $z$).
In writing these last two formulae, we have made use of the optical
theorem.

Similarly, we can write the cross-section for {\bf vector meson production}:
\begin{equation}
\left.\frac{d \sigma}{dt} \right|_{t=0} = \frac{1}{16 \pi} \left[ \int dz d^2r
\; \psi^*_V(z,r) \psi_{\gamma}(z,r) \sigma(z,r) \right]^2, \label{VM}
\end{equation}
where $\psi_V$ is the meson wavefunction.

For {\bf photon dissociation} processes, we want to sum incoherently over the
cross-sections to scatter into all possible final states, i.e. 
\begin{equation}
\left.\frac{d \sigma}{dt} \right|_{t=0} =  \frac{1}{16 \pi} \int dz d^2r
\; |\psi_{\gamma}(z,r)|^2 \sigma(s,r)^2.
\label{DD}
\end{equation}

So with nothing more than a bit of quantum mechanics and our definition
of diffraction we have arrived at these useful formulae. In particular
note that the elastic scattering amplitude and the photon dissociation
cross-section only involve the photon wavefunction (calculable in QED)
and the universal cross-section, $\sigma(s,r)$. The essential physics of
diffraction lives in $\sigma(s,r)$ (e.g. pomeron exchange, gluon ladders,...).
In order to proceed further, and gain some insight into the aforementioned
$W$-dependencies, we need to input a bit more physics. 

The photon wavefunction is calculated from the vacuum polarization graph and
possesses the following properties: (1) an exponential suppression sets in
for $r^2 > 1/[Q^2 \bar{z}+m^2]$; (2) $|\psi_{\gamma}^L|^2 \sim Q^2 \bar{z}^2$;
(3) $|\psi_{\gamma}^T|^2 \sim Q^2 \bar{z}$. The superscripts label the
mode of polarization, $\bar{z} \equiv z(1-z)$ and $m$ is the quark
mass. Gousset discussed the large size behaviour of the photon
wavefunction \cite{Gousset}. 

The dipole cross-section, $\sigma(s,r)$, must vanish in
proportion to $r^2$ as $r^2 \to 0$. This is the colour transparency
property which follows directly from QCD. For large $r$ we expect $\sigma(s,r)$
to saturate to some typical hadronic size, $R^2$ (due to confinement). We
are now ready to make some qualitative statements about photon induced
diffraction phenomena.    

We'll start with the diffraction dissociation cross-section, (\ref{DD}),
and look seperately at the contributions from large size $q \bar{q}$
pairs (i.e. $r > R$) and small size pairs (i.e. $r<1/Q$). For the large
size pairs the important range of the $z$ integral comes from the
end-points, where $\bar{z} < 1/[Q^2 R^2]$ (these are the only regions
which don't feel the exponential suppression from the tail of the
wavefunction), i.e. the $q \bar{q}$ pair is produced with a highly
asymmetric partitioning of the photon energy.
No such restriction is present for scattering small size pairs.
We can get a feel for what is going on without having to go into too
much detail. 

For the large size pairs we can write
\begin{equation}
\left.\frac{d \sigma}{dt} \right|_{t=0} \sim \frac{1}{Q^2 R^2} \cdot R^2
\cdot Q^2 \left(\frac{1}{Q^2 R^2}\right)^{n} \cdot R^4.
\end{equation}
The first factor on the rhs ($1/[Q^2 R^2]$) is from the volume of the
$z$ integral, the second ($R^2$) is from the $r$ integral, the third
is the wavefunction factor ($n=1$ for transverse photons and $n=2$ for
longitudinal photons) and the final factor is $\sigma^2 \sim R^4$. Thus
the rate induced by transverse photons is $\sim R^2/Q^2$ whilst that by
longitudinal photons is suppressed by an additional factor of $Q^2$,
i.e. $\sim 1/Q^4$. The additional factor of $\bar{z}$ in the
longitudinal photon wavefunction makes all the difference by suppressing
the $z$ end-point contribution. 

For the small size pairs, similar reasoning gives
\begin{equation}
\left.\frac{d \sigma}{dt} \right|_{t=0} \sim 1 \cdot \frac{1}{Q^2} \cdot Q^2
\cdot \frac{1}{Q^4} \sim \frac{1}{Q^4}.
\end{equation}
The $z$ volume gives the factor unity, the $r$ volume is now $\sim
1/Q^2$ and the photon wavefunction simply gives a factor $\sim Q^2$
(regardless of the polarization). The final factor is from 
$\sigma^2 \sim r^4 \sim 1/Q^4$. The contribution is
therefore higher twist.
  
We have arrived at the interesting
conclusion that {\it there is a leading twist contribution to the diffraction
dissociation rate and that it is a result of scattering large size $q
\bar{q}$ pairs produced by transversely polarized photons}. 
The HERA data support this picture, except perhaps for the fact that
the qualitative picture I've just presented suggests that $\alpha_P(0)
\approx 1.08$ should be observed (since the dominant contribution comes
from scattering large size $q \bar{q}$ pairs). 
The fact that a larger value is seen
is interesting and presumably arises because of QCD corrections which
build up an anomalous dimension which leads to an enhancement of the
short distance contribution. 

Now let's turn to vector meson production (\ref{VM}). For ``heavy'' mesons
(e.g. $J/\psi$), the non-relativistic approximation leads us to assume
that $| \psi_V|^2 \sim \delta(z-1/2)$ (or else the meson could not be
bound together). There is no end-point contribution and the quark
mass is large therefore the contribution from large size pairs is
exponentially suppressed. The rate for producing $J/\psi$ mesons off
on-shell photons rises rapidly with increasing $W$. This is a
characteristic of perturbation theory and is in accord with our
conclusion that only small size $q \bar{q}$ pairs need be considered.

For light mesons the situation is much more complicated and, not
surprisingly, depends critically on the end-point behaviour of the meson
wavefunction \cite{Brodsky}. 
For example, if we assume \footnote{This really is an
educated guess, it is a real challenge to understand the light meson
wavefunction.} that $\psi_V^* \psi_{\gamma}^T
\sim \bar{z}^{m+1/2}$ and that $\psi_V^* \psi_{\gamma}^L \sim
\bar{z}^{m+1}$ then it follows (following precisely the same reasoning
that led to the estimates for the dissociation rate) from (\ref{VM}) that
\begin{equation}
\frac{\sigma_T(r>R)}{\sigma_T(r<1/Q)} \sim (Q^2 R^2)^{1-2m}
\end{equation}
and
\begin{equation}
\frac{\sigma_L(r>R)}{\sigma_L(r<1/Q)} \sim (Q^2 R^2)^{-2m}.
\end{equation}     
Putting $n=1/2$ (which seems reasonable), means that the
production rate off transverse photons is sensitive to all sizes
(i.e. both perturbative and non-perturbative configurations) whilst the
rate off longitudinal photons is dominated by scattering of small size
pairs (perturbative). {\it Light meson production is thus a potentially very
interesting mix of soft and hard physics}. Information which will help
untangle what is going on comes in the form of measurements of 
$\sigma_L/\sigma_T$ and the variation of the total rate with $Q^2$ and $W$.  

\subsection{Pomeron parton densities}
Regge theory inspires the factorization of the structure function,
$F_2^{D(3)}$, extracted from high $Q^2$ photon dissociation
\cite{Arneodo,Ebert,Piotrzkowski}, i.e.
\begin{equation}
F_2^{D(3)}(\beta,Q^2,x_P) = f_P(x_P) F_2^P(\beta,Q^2) + \; {\rm
secondary} \; {\rm exchanges}.
\end{equation}
The data support this picture and are moving into the domain where they
can really test the notion of universal pomeron parton distribution
functions. At present, a model with DGLAP evolution describes the ZEUS data
on diffractive dijet production in photoproduction and on $F_2^{D(3)}$
\cite{Puga}. Also, H1 results on the hadronic final state (high $p_t$ 
particle production, energy flows and charm production in DIS dijets in
both DIS and photoproduction) are all consistent with the DGLAP approach
\cite{Ebert}. 

The universality of the pomeron parton densities is intimately connected to
the notion of the gap survival probability \cite{Engel}. Comparison between
data on direct and resolved processes, and from the Tevatron, will
certainly provide essential information in helping unravel the nature of
diffraction.  

\subsection{Squeezing the pomeron}
There are some rare diffractive processes whose rates can be calculated
purely using perturbative QCD. 

Hautmann presented results on the $\gamma^* \gamma^*$ total
cross-section \cite{Hautmann}. 
It is extracted (using the optical theorem) from the
elastic $\gamma^* \gamma^*$ amplitude at $t=0$, so is concerned with the
physics of diffraction. Since the photons are way off shell, they
scatter perturbatively via exchange of ``reggeized gluons'' between
their respective $q \bar{q}$ pairs. There are no hadrons to worry about,
so the calculation is very clean and worth looking for at LEP2 and
beyond.

An even better \footnote{There is no need to worry about diffusion
effects} way of keeping things perturbative is to look at high-$t$
diffraction. For example, one can look for a pair of high $p_t$ jets which are
separated by a big rapidity gap. Presumably one is looking at
parton-parton elastic scattering at $-t \approx p_t^2$ and, since
there's a gap, without exchange of colour. The fraction of dijet events
with a gap to all diject events as a function of increasing gap size has
been presented by ZEUS \cite{Hayes} out to gaps of $\approx 4$ units and
by D0 \cite{May} out to gaps of $\approx 6$ units. To really unravel the
important physics behind these data requires an understanding of the gap
survival probability. A very similar process that can be studied at HERA
and which avoids the issue of gap survival is high-$t$ vector meson
production (the proton dissociates to produce a forward jet, which,
since it need not be seen, means bigger gaps are admitted) \cite{jeff}. 
Both H1 and ZEUS are starting to accumulate good data on this process
\cite{Kohne,Crittenden}. 

\section{Jets: rates and shapes}

For an introduction to jet photoproduction, I refer to
Patrick Aurenche's presentation \cite{Aurenche}.

The structure of the virtual photon is starting to be examined at HERA,
and Rick presented results which showed that the $\gamma^*$ has a
significant ``resolved'' component for $Q^2$ as big as 50 GeV$^2$ in
those events where jets are produced with $E_T^2 > Q^2$ \cite{Rick}.  

\subsection{Dijets}

Data on two or more jets \cite{Hayes,Rick,Strickland} 
provides us with further options to test QCD and understand the nature
of the ``strongly interacting'' photon \cite{Aurenche}. ZEUS has defined 
direct enriched and resolved enriched samples by separating events
according to a cut at $x_{\gamma} = 0.75$. The direct enriched sample is
very sensitive to the small-$x$ gluon content of the proton: the more
backward the dijets, the lower the $x$ values in the proton that are
probed. Conversely, the resolved enriched sample is sensitive to the
gluon content in the photon. In addition, NLO calculations for the
dijet rates are now available for comparison with the data \cite{KK,O}.
Let's summarize the situation as it stands right now.

For $x_{\gamma} > 0.75$, the NLO theory does a good job \cite{Hayes}. 
However there remains quite a large contamination from the large-$x$
part of the photon quark distribution functions. This arises because of
the harder form of the photon quark densities. To unravel the effects of
the low-$x$ gluons in the proton from the large-$x$ quarks in the photon
requires a tighter cut on $x_{\gamma}$. To facilitate a clean comparison
between data and theory, the ZEUS collaboration has started to use the 
$k_t$-cluster algorithm \cite{ktclus}.

For $x_{\gamma} < 0.75$ the theory falls
well below the data for the lowest $E_T$ forward dijets \cite{Hayes}. The
effect exhibits a strong dependence upon the $E_T$ cut, which suggests
that it cannot be explained by modifying the parton distribution functions of
the photon in any sensible way. Presumably, this is the same problem
as that which has been encountered for the single jets \cite{1jet},
i.e. H1 and ZEUS both see an excess of single jet events in the forward
direction for $E_T < 15$ GeV (see \cite{Aurenche}). We really need to
understand what is going on before we can extract the gluon density of the
photon. Furthermore, these forward jets are fatter than might naively be
expected \cite{Strickland}.

A likely explanation for this effect could be due to the presence of a
large soft underlying event. Multiple parton interactions (MI) simulate 
(at least part of) this physics \cite{Engel,Chyla,Pancheri}. 
MI are anticipated on the grounds that 
forward jets at low $E_T$ are produced as a result of interactions
between slow partons in the colliding particles. We know that QCD
predicts a proliferation of these slow partons, and as such it may well
be that more than one pair of them can interact in each $\gamma p$ interaction.
MI can describe the broader nature of the forward jets
\cite{Strickland} and also increase the 
cross-sections for forward jet production, e.g. see \cite{BFS}.
One way of unambiguously identifying MI might be to look at higher (3 or
4) jet rates \cite{Strickland}. 

B\"urgin presented the OPAL results of jets in $\gamma \gamma$ \cite{Burgin}. 
Separation of events into classes involving direct and/or
resolved photons was performed via the $x_{\gamma}^{\pm}$ variable and
good agreement with the NLO calculations of Kleinwort \& Kramer
\cite{KlK} were
found. However, the error bars are still too large to allow much
discrimination between different parton distribution function
parameterizations. 

In conclusion, the dijets provide information which complements the
single jet measurements. The data are now reaching a high level of
precision, and comparison with NLO theory has revealed a number of pressing
issues. In particular, we need to understand better the forward jets and
use the most appropriate jet algorithm. Once these issues have been
addressed, we can expect to gain further insight into the gluon content
of both the photon and proton.  

\subsection{Soft gluons}
The ZEUS dijet measurements have been made with a cut on the
minimum $E_T$ of the jets and the cut is the same for both jets. This
introduces a further theoretical problem. This arises because most of
the jets will be produced around the minimum allowable $E_T$, i.e. the
typical difference between the jet transverse momenta, $\Delta p_T$, will
be small. So, the 3 parton final state (which is present in the NLO 
calculation) must have
one of the partons either collinear with another, or very soft. The
collinear configuration is easy to deal with (it is factorized) but the
soft parton emission leads to a $\ln \Delta p_T$ contribution. This
large logarithm signals that multiple soft parton emission is
important. These soft parton effects can be
studied by looking explicitly at the $\Delta p_T$ distribution of the
dijets or they can be avoided by making a cut which keeps away from
$\Delta p_T \approx 0$ \cite{Aurenche}. 

Similar effects need to be considered in double prompt photon production
which is being observed at the Tevatron \cite{Chen}. D0 cut on the
photon transverse energies, i.e. $E_{T1}>14$ GeV and $E_{T2} > 13$
GeV. The need to sum the soft gluon effects can be seen in that the
theory overshoots data for $\Delta p_T < 3$ GeV.

\section{Prompt Photon Production}
The rates for prompt photon production seen by D0 and CDF suggest a
possible excess of events at low $E_T/\surd s$ \cite{Chen}. However,
Andreas Vogt pointed out that the data are within the theoretical
uncertainty.

Gordon presented results on prompt $\gamma$ plus jet at HERA
\cite{Gordon}. The NLO
calculation is coded into Monte Carlo. This process is sensitive to the
gluon density of the photon (for low $x_{\gamma}$ where there are no
data yet) and isolation cuts kill off the fragmentation contribution. We
saw the good agreement between these NLO calculations and the ZEUS data
\cite{Vaiciulis}. At present the data are statistics limited.

\section{Open Charm}
The photoproduction of charm quarks at large $p_T$ is a process which
involves two large scales, $p_T, m_c \gg \Lambda$ and as such, makes
life more complicated from the theoretical point of view. Good data,
which can be expected in the future (especially if the charm can be
tagged using a microvertex detector), will surely shed light on this
intriguing area. At present, there are two main routes used in
theoretical calculations. 

{\bf Massive charm:}
The charm quark mass is considered to provide the hard scale, as such
charm only ever appears in the hard subprocess and there is no notion of
radiatively generated charm in the sense of parton evolution. This means
that terms $\sim \alpha_s \ln(p_T^2/m_c^2)$ are not summed to all orders
(in the parton distribution and fragmentation functions). As such, we
might expect this approach to become less accurate when $p_T^2 \gg
m_c^2$. However, it does provide a systematic way of accounting for
charm quark mass effects, which will be important for $m_c \sim p_T$.

{\bf Massless charm:}
In this approach, the charm quark is treated as massless (above
threshold), and as such is treated like any other light quark in the
parton evolution equations and hard subprocess cross-sections. 
The $\ln(p_T/m_c)$ terms are now summed to all orders, but charm quark
mass effects are ignored. So, this approach should get better as
$p_T/m_c$ increases.

Gladilin presented new results from ZEUS \cite{Gladilin}. At present the
data lie in the intermediate region where $p_T < 10$ GeV, i.e. 
it is not clear which, if any, of the two approaches should be
used. In order to compute the inclusive $D^*$ rate, one needs the
appropriate fragmentation function. Either the Peterson \cite{Peterson}
form or the $x^a (1-x)^b$ form do a good job, and can be well
constrained by $e^+ e^-$ data.

Initial comparisons between theory and data suggest that the massive
charm calculation is too low, e.g see \cite{Gladilin}. 
However, the full NLO calculations require that the
fragmentation function be consistently extracted from the $e^+ e^-$
data. When this is done, Cacciari and Greco find that the theoretical
predictions are increased significantly relative to what is found using
the softer (LO) fragmentation functions \cite{Aurenche,CG}. 
This is true for both massless and massive charm calculations and, within 
theoretical uncertainties, both are now consistent with the present data. 

More data at large $p_T$ and increased statistics at intermediate $p_T$
will certainly help in our study of the interplay of $\ln(p_T^2/m_c^2)$
and $m_c^2/p_T^2$ effects. In addition, for $p_T \gg m_c$ we have the
possibility to study the ``intrinsic'' charm within the photon   
(charm in dijets offers good prospects here).  

In $e^+ e^-$, Andreev presented new results from L3
\cite{Andreev}. The open charm total cross-section agrees with the NLO
calculation of Drees et al. \cite{Drees}. 
We can look forward to more statistics which
will allow comparison with differential distributions.

\section{Charmonium}

\begin{figure} 
\centerline{\epsfig{file=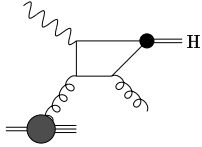,height=0.8in,%
bbllx=169pt,bblly=388pt,bburx=286pt,bbury=475pt}}
\caption{Leading contribution to quarkonium photoproduction.}
\label{LOCSM}
\end{figure}

Originally, inelastic photoproduction of charmonium,
e.g. $J/\psi$, was advertised as an ideal way to extract the gluon density
in the proton (since it is driven by photon-gluon fusion into a $Q
\bar{Q}$ pair). More recently, NLO calculations have put a dampener on this
goal \cite{CnK}. However, there has been a great deal of
recent interest in the non-relativistic QCD (NRQCD) approach to
heavy quarkonium production, and the inelastic photoproduction of heavy
quarkonia provides the ideal opportunity to test NRQCD. 

Bodwin, Braaten and Lepage derived a factorization formula which
describes the inclusive production (and decay) of a heavy quarkonium state
\cite{BBL}. In the case of photoproduction, Fig.\ref{LOCSM} shows the
lowest order contribution. 
The NRQCD factorization formula for the corresponding
cross-section reads
\begin{equation}
d\sigma(H+X) = \sum_{n} d\hat{\sigma}(Q \bar{Q}[n]+X) \langle O^H_n
\rangle. 
\end{equation}
$X$ denotes that the process is inclusive, $d\hat{\sigma}(Q
\bar{Q}[n]+X)$ is the perturbatively calculable cross-section for
$\gamma p \to Q \bar{Q} + X$ and it can be written as a series expansion
in $\alpha_s(m_Q)$. The $Q \bar{Q}$ pair is produced with
quantum numbers $n$. The matrix element, $\langle O^H_n \rangle$,
contains the long distance physics associated with the formation of the
quarkonium state $H$ from the $Q \bar{Q}$ state -- it is
essentially the probability that the pointlike $Q \bar{Q}$ pair forms
$H$ inclusively. The typical scale
associated with this part of the process is $\sim m_Q v$ which is much
smaller than $m_Q$ ($v$ is the relative velocity of the $Q \bar{Q}$
pair, and is small for heavy enough quarks). This hierarchy of scales
underlies the NRQCD factorization. Note that the $Q \bar{Q}$ state is
not restricted to having the same quantum numbers as the
meson. Fortunately, there exist ``velocity scaling rules'' which allow us
to identify which states, $n$, are the most important. More precisely,
the ``velocity scaling rules'' order the operators $\langle O^H_n
\rangle$ according to how many powers of $v$ they contain,
i.e. relativistic corrections can be computed systematically. 

The NRQCD approach therefore provides us with a systematic way of
computing inclusive heavy quarkonium production (modulo corrections
which are suppressed by powers of $\sim \Lambda/m_Q$). The strategy is
first to organise the sum over $n$ into an expansion in $v$ and then to
systematically compute $d \hat{\sigma}$ order-by-order in
$\alpha_s(m_Q)$. Technically, we do not a priori know where our efforts
are best placed, i.e. do we work at lowest order in $v$ and to NLO in
$\alpha_s$ or do we attempt to work at higher orders in $v$, but
computing each hard subprocess to lowest order? We need to know $v$ in
order to judge better what to do.

One final word before moving on to discuss $J/\psi$ photoproduction. 
For small $p_T$, NRQCD factorization is likely to break down, due to
contamination from higher twist effects. Also, one expects breakdown of
the NRQCD approach in the vicinity of the elastic scattering region,
i.e. $z \to 1$ where $z$ is the fraction of the photon energy carried by the
quarkonium (see later). 

Inelastic photoproduction of $J/\psi$ is something which has already
been measured at HERA. Let's see how the theory shapes up. To lowest
order in the velocity expansion, $[n] = [1,^3 \! S_1]$. The first entry in
the square brackets tells us that the $c \bar c$ is in a colour singlet
state, whilst the second entry tells us the spin and angular momentum of
the state. Not surprisingly, to lowest order in the velocity expansion,
the $c \bar{c}$ must be produced with the same quantum numbers as the
$J/\psi$. This is just the colour singlet model (CSM) of old. The lowest
order diagram which can contribute is shown in
Fig.\ref{LOCSM} and
$$ \langle O^{J/\psi}[1,^3 \! S_1] \rangle \sim | \phi(0) |^2 $$
where $\phi(0)$ is the wavefunction at the origin (it can be extracted
from the electronic width of the $J/\psi$). NLO$(\alpha_s)$ corrections
have been computed \cite{ZSZK} and shown to be large.
The NLO corrections enhance the LO calculation and lead to a reduced 
sensitivity to the gluon density in the proton.

\begin{figure}[!h] 
\centerline{\epsfig{file=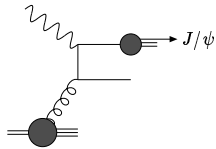,height=0.8in,%
bbllx=169pt,bblly=388pt,bburx=296pt,bbury=475pt}}
\caption{Fragmentation contribution to the production of the $J/\psi$.}
\label{frag}
\end{figure}

One might well ask as to the significance of the resolved photon
contribution. It is important at small enough $z$ \cite{CnK,GRS,KnK}.
In addition, for mesons produced at high enough $p_T$ we have an
additional scale to consider and terms which are leading in $\alpha_s$
can be suppressed by powers of $\sim m_c^2/p_T^2$. This is
true for example of the diagram shown in Fig.\ref{LOCSM} relative
to that shown in Fig.\ref{frag}. 
The latter fragmentation contribution is higher order in
$\alpha_s$, however there is one less hard quark propagator and so it
will dominate for large enough $p_T$. Fragmentation contributions and
resolved photon contributions are not important in computing the total
rate for $z > 0.4$ (which is essentially where the data are).

Going to NLO in the velocity expansion means moving away from the
CSM. For the first time we encounter colour octet contributions. In
particular, the LO$(\alpha_s)$ diagram is again that of
Fig.\ref{LOCSM} \footnote{There is a lower order contribution in which 
no gluon is
radiated off, however this would give a contribution only at $z=1$ which
lies outside our region of interest.}
now the $c \bar{c}$ can be formed in one of 5 states, i.e.
$[n] = [8,^1 \! S_0], [8,^3 \! S_1], [8,^3 \! P_{0,1,2}]$. 
The price one pays for
having to convert this state into the $J/\psi$ is an extra power of
$v^4$ relative to the colour singlet matrix element, i.e.
$$ \langle O^{J/\psi}[n] \rangle  \sim v^4 \langle O^{J/\psi}[1,^3 \! S_1]
\rangle.$$ It turns out that the $v^4 \sim 0.01$ suppression of the long
distance matrix elements is partially compensated for by a strong
enhancement of
the corresponding short distance cross-section. In particular, this is
so for the $[8,^1 \! S_0]$ and $[8,^3 \! P_{0,2}]$ states. The colour singlet
matrix element can be extracted from data, e.g. the leptonic width of the
$J/\psi$. Likewise, we need to fit these new matrix elements to data (or
extract them from lattice calculations). It
is therefore clear that a test of the NRQD framework requires
data from different sources -- the challenge being to find a consistent
description. This is a particularly topical issue, since an explanation
of the Tevatron excess of direct $J/\psi$ and $\psi'$ production needs, in
addition to fragmentation contributions, colour octet contributions
\cite{BF}. One
can use the Tevatron data to fit the relevant matrix elements. The validity
of this explanation can be checked on comparing to data
which can be obtained from HERA. Unfortunately, the matrix elements
which are important at the Tevatron are not so important in $J/\psi$
photoproduction for $z > 0.4$.  
However, the key matrix element for the Tevatron ($[8,^3S_1]$) does play
a key role in the region of lower $z$, where (for large enough $p_T$) the
dominant contribution comes from the fragmentation mechanism via
resolved photons \cite{CnK,KnK}.    
Another process which is sensitive to the $[8,^3S_1]$ state is
the photoproduction of $J/\psi + \gamma$ (where the photon is produced
in the hard subprocess, i.e. not via the radiative decay of a $P$-wave
quarkonium) \cite{CGK}. So, with the anticipated increase in statistics, we
can really expect to test NRQCD at HERA. 
Going back to the $J/\psi$, there are some weak constraints
on the important matrix elements from the Tevatron data and these have been
used in the theoretical calculations of \cite{Mike}. The HERA
data on the $z$ distribution compare very well
with the colour singlet calculation \cite{Aurenche,Gladilin}. 
The colour octet contribution however, is much too large at large $z$. 
Thus the HERA data is not supporting a large colour octet
contribution at large $z$. However, one must be careful in interpreting this as
evidence against the NRQCD approach, since the $z \to 1$ region is
sensitive to higher order non-perturbative contributions which lead to
the breakdown of the NRQCD expansion \cite{BRW}.

In addition to the processes just discussed, increased statistics will
allow measurement of other meson states, e.g. $\psi'$ and $\Upsilon$,
which will certainly further test our understanding of QCD. 

\section{Outlook}
We really need to improve our understanding of the $e \gamma$ final
state \cite{DJM} if we are to reduce the systematic uncertainty which
presently dominates the experimental measurements of
$F_2^{\gamma}$. 

An improved understanding of the soft underlying event and of
multiple interactions is needed in order to understand
the gap survival probability in diffractive events.
It is also needed for a better understanding of forward jets at HERA 
(which can then be used to extract the gluon density of the photon).

We can look forward to the accumulation of data on prompt photon
production, $\gamma^* \gamma^*$ reactions and virtual photon structure, 
high $t$ diffraction and diffractive meson production at high $Q^2$.
Comparison of diffraction data from deep inelastic scattering with that
from photo- (and hadro-) production will play a central role in
developing our understanding of diffraction. 
It would be great to see data on diffraction in $\gamma \gamma$
collisions \cite{Engel}. Meanwhile, the search for the
odderon will continue \cite{Tapprogge}.

Charm production will provide tests of NRQCD and allow us to unravel the
subtle issues associated with open charm production.

\section*{Acknowledgements}
Thanks to the organizers for such an enjoyable conference. I also want
specifically to thank David Miller and Lionel Gordon for helping me put
the talk together.

\end{document}